\documentclass[proof]{WileyASNA-v1}
\usepackage{slashed}
\usepackage{dsfont}
\usepackage{braket}
\articletype{Article Type} 

\received{1 November 2022}
\revised{3 November 2022}
\accepted{4 November 2022}
\begin{document}

\title{Quark matter phase diagram under the influence of strong magnetic fields with a nonlocal chiral model}

\author[1,2]{J. P. Carlomagno}
\author[3]{S. A. Ferraris}
\author[1,2]{D. Gómez Dumm}
\author[2,3]{A. G. Grunfeld}
\authormark{A.G.Grunfeld*}

\address[1]{\orgdiv{IFLP, CONICET departamento de Física}, \orgname{Facultad de Ciencias Exactas, Universidad Nacional de La PLata}, \orgaddress{C.C. 67, (1900) \state{La Plata}, \country{Argentina}}}

\address[2]{\orgname{CONICET}, \orgaddress{Rivadavia 1917, (1033),\state{Buenos Aires}, \country{Argentina} }}

\address[3]{\orgdiv{Physics Department}, 
\orgname{Comisión Nacional de Energía Atómica (C.N.E.A)}, \orgaddress{Avenida del Libertador 8250 (1429), \state{Buenos Aires}, \country{Argentina}}}

\corres{*Av. Gral. Paz 1499, San Martin, Buenos Aires.  \email{grunfeld@tandar.cnea.gov.ar}}

\abstract{We study the phase diagram in the $T-\mu$ plane for quark matter under the influence of a strong uniform magnetic field $\vec{B}$, in the framework of a non-local extension of the two-flavor Polyakov–Nambu–Jona-Lasinio model. We analyze the deconfinement and chiral symmetry restoration transitions in the mean field approximation. For the considered parameterization, it is found that there is always a critical end point (CEP) in the $T-\mu$ plane that separates a first-order transition line from a smooth crossover. The location of the CEP is studied as a function of the magnetic field.}

\maketitle

\section{Introduction}
The study of the phase diagram of QCD ---which involves features of a non-perturbative regime--- is an interesting subject, as it deals with a behavior of strongly interacting matter that has applications in various fields such as cosmology, heavy ion collisions and compact stars physics. Moreover, it is interesting to analyze the structure of the phases in the presence of strong magnetic fields. In fact, extremely high magnetic fields ($10^{16}$ T, $eB \simeq 1.5$ GeV$^2$) could have been present during the cosmological electroweak phase transition \citep{Vachaspati:1991nm}. Also, in ultra-peripheral heavy-ion collisions, the generated magnetic fields are proportional to the collision energy, which reaches $4 \times 10^{14}$ T \citep{Deng:2012pc}. In an astrophysical scenario, the magnetic field on the surface of magnetars is expected to be of the order of $10^{10}$ T \citep{1992ApJ...392L...9D}.  

In the present work, we study the phase diagram at finite temperature and density of quark matter under the influence of a strong magnetic field. Our model is based on a non-local extension of the two-flavor Nambu-Jona-Lasinio model \citep{dumm2006covariant} together with the inclusion of a Polyakov-loop potential, which mimics confinement and leads to a critical temperature for deconfinement and chiral symmetry restoration compatible with lattice QCD (LQCD) calculations. For different strengths of the magnetic field, we focus on the position in the $T-\mu$ plane of the critical endpoint (CEP) that separates first order and crossover-like chiral restoration transition lines. The present work is a complementary analysis of previous works, Refs.~\citep{dumm2017strong} and  \citep{Ferraris_2021}, where finite temperature and finite density were considered separately.  

\section{formalism}
Let's start by defining the Euclidean action in our non-local NJL model for two quark flavors $u$ and $d$,
\begin{equation}
S_{E}=\int d^{4}x \left[\bar{\psi}(x)\left(-i\slashed{\partial} + m_{c}\right)\psi(x)-\frac{G}{2}j_{a}(x)j_{a}(x)\right],  
\label{accionE}
\end{equation}
where $\psi$ stands for the fermionic fields, $G$ is a coupling constant, and we assume that the current quark mass $m_{c}$ is the same for both flavors, $m_{c}=m_{u}=m_{d}$. The non-local currents $j_{a}(x)$ are defined as           
\begin{equation}
j_{a}(x)=\int d^{4}z\ \mathcal{G}(z)\bar{\psi}\left(x + \frac{z}{2}\right)\Gamma_{a}\psi\left(x - \frac{z}{2}\right), \label{corrientes}
\end{equation}
where $\Gamma_{a}=(\mathds{1},i\gamma_{5}\vec{\tau})$ and $\mathcal{G}(z)$ is a nonlocal form factor. In order to introduce the interaction with an external magnetic field $\vec{B}$ one has to replace the partial derivative $\partial_{\mu}$ in the kinetic term of the effective Euclidean action Eq.~(\ref{accionE}) by the covariant derivative 
\begin{equation}
\partial_{\mu} \rightarrow D_{\mu}\equiv \partial_{\mu}-i\hat{Q}\mathcal{A}_{\mu},         
\end{equation}
where $\mathcal{A}_{\mu}$ is an external electromagnetic gauge field, and $\hat{Q}={\rm diag}(q_{u}, q_{d})$, with $q_{u}=2e/3$, $q_{d}=-e/3$, is the electromagnetic quark charge operator. This replacement also implies a change in the non-local currents Eq.~(\ref{corrientes}) given by \cite{noguera2008nonlocal,gomez2011pion,dumm2006covariant} 
\begin{gather}
\psi(x-z/2) \rightarrow \mathcal{W}(x,x-z/2)\psi(x-z/2),\nonumber\\
\psi(x+z/2)^{\dagger} \rightarrow \psi(x+z/2)^{\dagger}\mathcal{W}(x+z/2,x),
\end{gather}
where the function $\mathcal{W}(s,t)$ is defined as 
\begin{equation}
\mathcal{W}(s,t)=P\,\exp\left[-i\int_{s}^{t} dr_{\mu}\hat{Q}\mathcal{A}_{\mu}(r) \right].     
\label{Funcw}
\end{equation}
 For simplicity, we take here a straight line path connecting $s$ with $t$ in the integral of Eq.~(\ref{Funcw}). This ansatz was originally proposed in Ref.~\citep{bloch1952field} and is commonly used in the literature.
 
 To proceed, we consider a constant and homogeneous magnetic field oriented along the 3-axis and work in the Landau gauge, in which we have $\mathcal{A}_{\mu}=Bx_{1}\delta_{\mu 2}$. Under this gauge choice, the function $\mathcal{W}(s,t)$ defined  in Eq.~(\ref{Funcw}) is given by
 \begin{equation}
\mathcal{W}(s,t)=\exp\left[-\frac{i}{2}\hat{Q}B\left(s_{1} + t_{1}\right)\left(t_{2} - s_{2}\right) \right].     
\label{Funcw2}
\end{equation}
Since the degrees of freedom of quark fields are not observed at low energies, the fermions can be integrated out, writing the action in terms of scalar and pseudo-scalar fields $\sigma(x)$ and $\vec{\pi}(x)$, respectively. The bosonized action is given by \cite{noguera2008nonlocal,gomez2011pion} 
\begin{equation}
S_{\rm bos}=-\ln\,\det\mathcal{D}_{x,x'} + \frac{1}{2G}\int d^{4}x \left[\sigma(x)\sigma(x)+\vec{\pi}(x)\cdot\vec{\pi}(x) \right],
\label{Sbos}   
\end{equation}
with
\begin{align}
\mathcal{D}_{x,x'} = &~ \delta^{(4)}(x-x')\left(-i\slashed{D}+m_{c}\right) + \mathcal{G}(x-x')\gamma_{0}\nonumber\\
&\times\mathcal{W}(x,\bar{x})\gamma_{0}\left[\sigma(\bar{x}) + i\gamma_{5}\vec{\tau}\cdot\vec{\pi}(\bar{x})\right]\mathcal{W}(\bar{x},x'), 
\end{align} 
where $\bar{x}=(x+x')/2$ for the neutral mesons.        

In models in which spontaneous symmetry breaking occurs, as in our case, mesonic fields can be written in terms of the corresponding vacuum expectation values and their fluctuations, $\sigma(x)=\bar{\sigma} + \delta \sigma(x)$ and $\vec{\pi}(x)=\delta \vec{\pi}(x)$. Here we work in the mean field approximation (MFA) where the field $\sigma(x)$ has a non-trivial translational invariant mean field value $\bar{\sigma}$, while for reasons of symmetry the mean field value of pseudoscalar field is $\vec{\pi}(x)=\vec{0}$. After some calculations one gets the MFA bosonized action per unit volume   
\begin{equation}
\frac{S_{\rm bos}^{\rm MFA}}{V^{(4)}}=\frac{\bar{\sigma}^{2}}{2G} - \frac{N_{c}}{V^{(4)}}\sum_{f=u,d} {\rm tr}\,\ln\left[\mathcal{D}_{x,x'}^{{\rm MFA},f}\right], 
\label{S_bos_MFA_Vol}
\end{equation}
where $N_{c}$ is the number of colors. To calculate the traces over Dirac and coordinate spaces, it is convenient to perform a Ritus transform of $\mathcal{D}_{x,x'}^{{\rm MFA},f}$ \citep{ritus1978method}.      

Next, using the standard Matsubara formalism we include in our model both finite temperature $T$ and chemical potential $\mu$, based in previous works Refs.~\citep{dumm2017strong} and \citep{Ferraris_2021} where $T$ and $\mu$ were considered separately. The purpose of this article is to study the combined effects of both thermodynamic variables in the system. Now, to account for the confinement/deconfinement effects we also include the coupling of fermions to the Polyakov loop $\Phi$. We assume that quarks move in a constant color background field $\phi=ig\delta_{\mu0}G_{a}^{\mu}\lambda^{a}/2$, where $G_{a}^{\mu}$ are color gauge fields. It is convenient work in the so-called Polyakov gauge, in which the matrix $\phi$ is given in a diagonal representation $\phi = \phi_{3}\lambda_{3} +\phi_{8}\lambda_{8}$ with only two independent variables, $\phi_{3}$ and $\phi_{8}$. The trace of Polyakov loop $\Phi=\frac{1}{3}{\rm Tr}\,\exp(i\phi/T)$ is used as an order parameter for the confinement/deconfinement transition. This parameter is expected to be real owing to the charge conjugation properties of the QCD Lagrangian \citep{dumitru2005dense}; this implies $\phi_{8}=0$, and therefore $\Phi=[1+2\cos(\phi_{3}/T)]/3$. Finally, to describe the gauge field self-interactions we also include in the Lagrangian an effective potential $\mathcal{U}(\Phi,T)$. The expression used in this work is based on a Ginzburg-Landau ansatz \citep{ratti2006phases,scavenius2002k}
\begin{equation}
\frac{\mathcal{U}(\Phi,T) }{T^{4}} = \frac{-b_{2}(T)}{2}\Phi^{2} - \frac{b_{3}}{3}\Phi^{3}+\frac{b_{4}}{4}\Phi^{4},     
\end{equation}
where 
\begin{equation}
b_{2}(T) = a_{0}+a_{1}\left(\frac{T_{0}}{T} \right) + a_{2}\left(\frac{T_{0}}{T}\right)^{2} + a_{3}\left(\frac{T_{0}}{T} \right)^{3}.     
\end{equation}
with
\begin{center}
\begin{tabular}{  l  c  r  } 
 $a_{0}=6.75$, & $a_{1}=-1.95$, & $a_{2}=2.625$, \\ 
 $a_{3}=-7.44$, & $b_{3}=0.75$, & $b_{4}=7.5$. \\ 
\end{tabular}
\end{center}
The numerical values of the parameters were taken from Ref.~\citep{ratti2006phases} and we set the value $T_{0}=210$~MeV for light dynamical quarks.

Under these conditions, we can define the grand canonical thermodynamic potential at finite temperature $T$ and chemical potential $\mu$ of the system, under the influence of an external and homogeneous magnetic field, as      
\begin{align}
\Omega_{B,T,\mu}^{\rm MFA}&=\frac{\bar{\sigma}^{2}}{2G}- T\sum_{n=-\infty}^{\infty}\sum_{c=r,g,b} \sum_{f=u,d}\frac{|q_{f}B|}{2\pi}\int\frac{dp_{3}}{2\pi}\nonumber\\    
& \times\left[\ln\left(p_{||}^{2} + \left( M_{0,p_{||}}^{\lambda_{f},f} \right)^{2} \right)+\sum_{k=1}^{\infty}\ln~\Delta_{k,p_{||}}^{f} \right] + \mathcal{U}(\Phi,T), 
\label{granpotMFA}
\end{align}
where
\begin{align}
\Delta_{k,p_{||}}^{f}=&~\left( 2k|q_{f}B|+p_{||}^{2} + M_{k,p_{||}}^{+,f}M_{k,p_{||}}^{-,f}\right)^{2}\nonumber\\    
& +p_{||}^{2}\left(M_{k,p_{||}}^{+,f} -M_{k,p_{||}}^{-,f}\right)^{2}
\end{align}
with 
\begin{equation}
M_{k,p_{||}}^{\lambda,f} = (1-\delta_{k,-1})m_{c} + \bar{\sigma}~g_{k,p_{||}}^{\lambda,f}.      
\label{masa_const}
\end{equation}
The expression in Eq.~(\ref{masa_const}) can be interpreted as a constituent quark mass in the presence of an external magnetic field. Here we used the definitions $k_{\pm}=k-1/2\pm s_{f}/2$, where $k$ stand for Landau levels, $s_{f}={\rm sign}(q_{f}B)$, and $p_{||}=[p_{3},(\omega_{n}-i\mu-\phi_{c})]$, $\omega_{n}=(2n+1)\pi T$ being Matsubara frequencies corresponding to fermionic modes. The subscripts $c=r,g,b$ and $f=u,d$ stand for color and flavor, respectively. Color background fields are given by $\phi_{r}=-\phi_{g}=\phi_{3}$ and $\phi_{b}=0$. 

The integral in Eq.~(\ref{granpotMFA}) is divergent and has to be regularized. We use a prescription similar to that used in Ref.~\citep{dumm2005characteristics}, namely                
\begin{equation}
\Omega_{B,T,\mu}^{\rm MFA, reg}=\Omega_{B,T,\mu}^{\rm MFA} - \Omega_{B,T,\mu}^{\rm free} + \Omega_{B,T,\mu}^{\rm free,reg}.    \end{equation}  
Here, $\Omega_{B,T,\mu}^{\rm free}$ is evaluated at $\bar{\sigma}=0$, keeping the interaction with the magnetic field and the Polyakov loop. The term corresponding to $\Omega_{B,T,\mu}^{\rm free,reg}$ can be written as 
\begin{align}
&\Omega_{B,T,\mu}^{\rm free,reg}=-\frac{3}{2\pi^{2}}\sum_{f}\left(q_{f}B\right)^{2}\left[\zeta^{'}(-1,x_{f}) + F(x_{f}) \right]\nonumber\\
&-T\sum_{c,f,k} \frac{|q_f B|}{2\pi}~\alpha_k \int \frac{dp}{2\pi}~ G_{k,p}^{f}(\phi_{c},\mu,T) 
\end{align}
with
\begin{align}
&F(x_{f}) =\nonumber \frac{x_{f}^{2}}{4}-\frac{1}{2}\left(x_{f}^{2} - x_{f}\right)\ln(x_{f})\\
&\nonumber G_{k,p}^{f}(\phi_{c},\mu,T)=\sum_{s=\pm} \ln \left\{ 1 + \exp[-( \epsilon_{kp}^f +i \phi_{c} + s \mu) /T ]\right\},    
\end{align}
where $x_{f}=m_{c}^{2}/(2|q_{f}B|)$, $\alpha_{k}=2-\delta_{k,0}$, $\epsilon_{kp}^{f}=\left( (2k|q_{f}B|) +p^{2} +m_{c}^{2}\right)^{1/2}$ and $\zeta^{'}(-1,x_{f})=d\zeta(z,x_{f})/dz|_{z=-1}$, where $\zeta(z,x_{f})$ is the Hurwitz zeta function. 

Finally, $\bar{\sigma}(B,T,\mu)$ and $\Phi (B,T,\mu)$ are obtained by solving the system of two coupled equations that minimize $\Omega_{B,T,\mu}^{\rm MFA, reg}$, viz.   
\begin{align}
\frac{\partial \Omega_{B,T,\mu}^{\rm MFA, reg}}{\partial \bar{\sigma}} = 0,    & \qquad    \frac{\partial \Omega_{B,T,\mu}^{\rm MFA, reg}}{\partial \Phi} = 0.     
\label{gap}
\end{align}
Note that there are
regions for which there is more than one solution for
each value of $T$ and $\mu$ (for fixed $B$). In that case we consider that the stable solution is the one corresponding to the overall minimum of the potential.
Given the thermodynamic potential, the expressions for all other relevant quantities can be easily derived. An important magnitude to be considered is the quark-antiquark condensate for each flavor, which is defined as
\begin{equation}
\braket{\bar{q}_{f}q_{f}}_{B,T,\mu}^{\rm reg}=\frac{\partial \Omega_{B,T,\mu}^{\rm MFA, reg}}{\partial \bar{m_{c}}}.      
\end{equation}

Finally, to determine the
characteristics of the chiral phase transition, in next section we will define the chiral susceptibility.

\section{numerical results}
We considered a Gaussian form factor to describe the non-local interactions,
\begin{equation}
g(p^{2})= exp\left( -p_{||}^{2}/\Lambda^{2} \right).     
\end{equation}
With this particular choice, the expression for the constituent quark masses takes the form    
\begin{align}
M_{k,p_{||}}^{\lambda,f} = ~&(1-\delta_{k,-1})~m_{c}\\     
&\nonumber + \bar{\sigma}\frac{\left( 1-|q_{f}B|/\Lambda^{2}  \right)^{k_{\lambda}}}{\left(1+|q_{f}B|/\Lambda^{2} \right)^{k_{\lambda}+1}}~exp\left(-p_{||}^{2}/\Lambda^{2}\right). 
\end{align}
In order to carry out our calculations, we need to specify the input parameters of the model, viz.\ $m_{c}$, $\Lambda$ and $g=G\Lambda^{2}$. Here we use the values $m_{c}=6.5$ MeV, $\Lambda=678$ MeV and $G\Lambda^{2}=23.66$ \citep{dumm2017strong}. They were obtained by fixing the empirical values $m_{\pi}=139$ MeV and $f_{\pi}=92.4$ MeV and (at $\mu=T=eB=0$) for pion mass and weak decay constant, respectively, and a phenomenologically reasonable
value of chiral quark condensate $-\braket{\bar{q}q}^{1/3}=230$ MeV.  

\begin{figure}[htb]
\begin{tabular}{c}
\vspace{-1.3cm}
\includegraphics[scale=0.31]{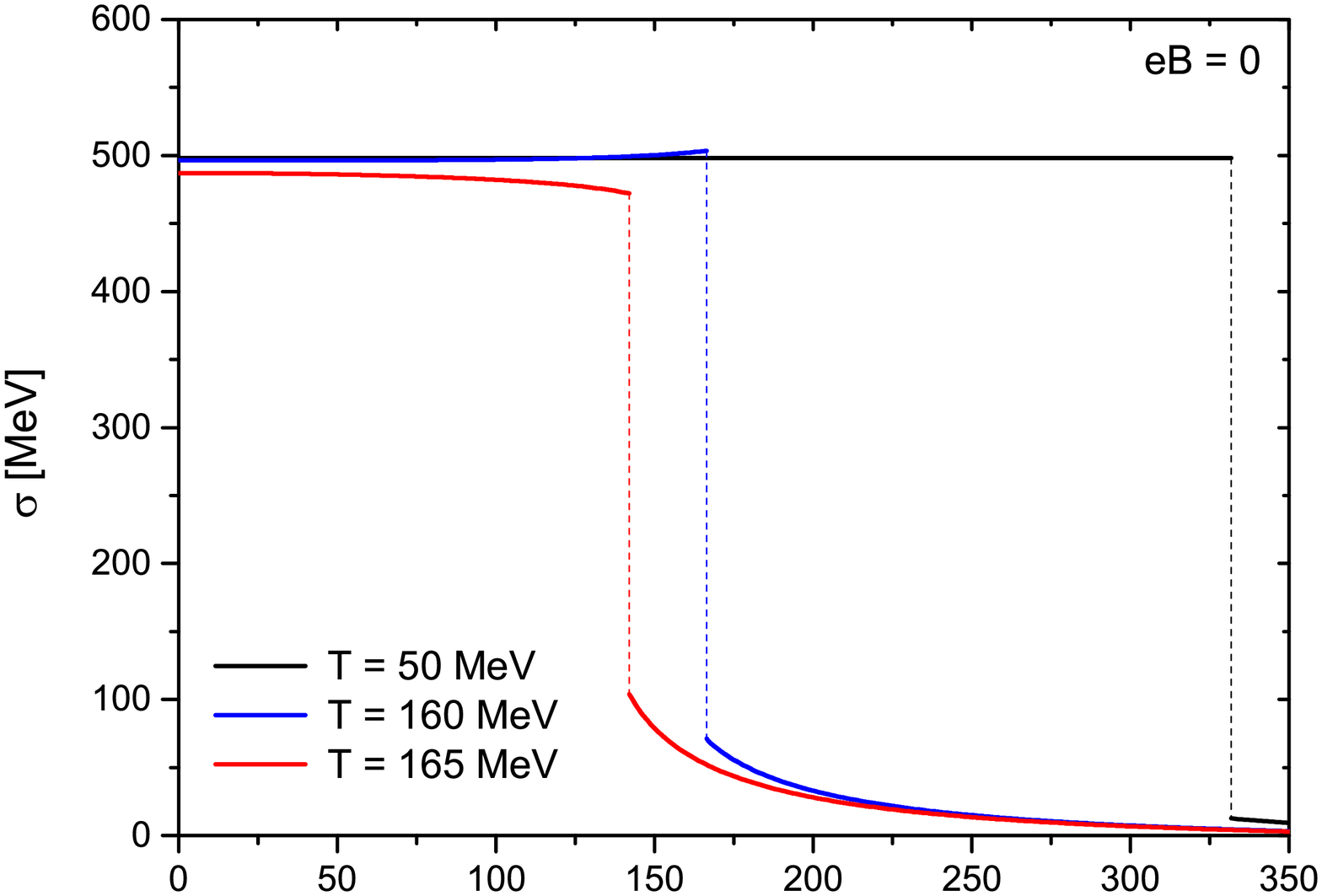}\\
\includegraphics[scale=0.31]{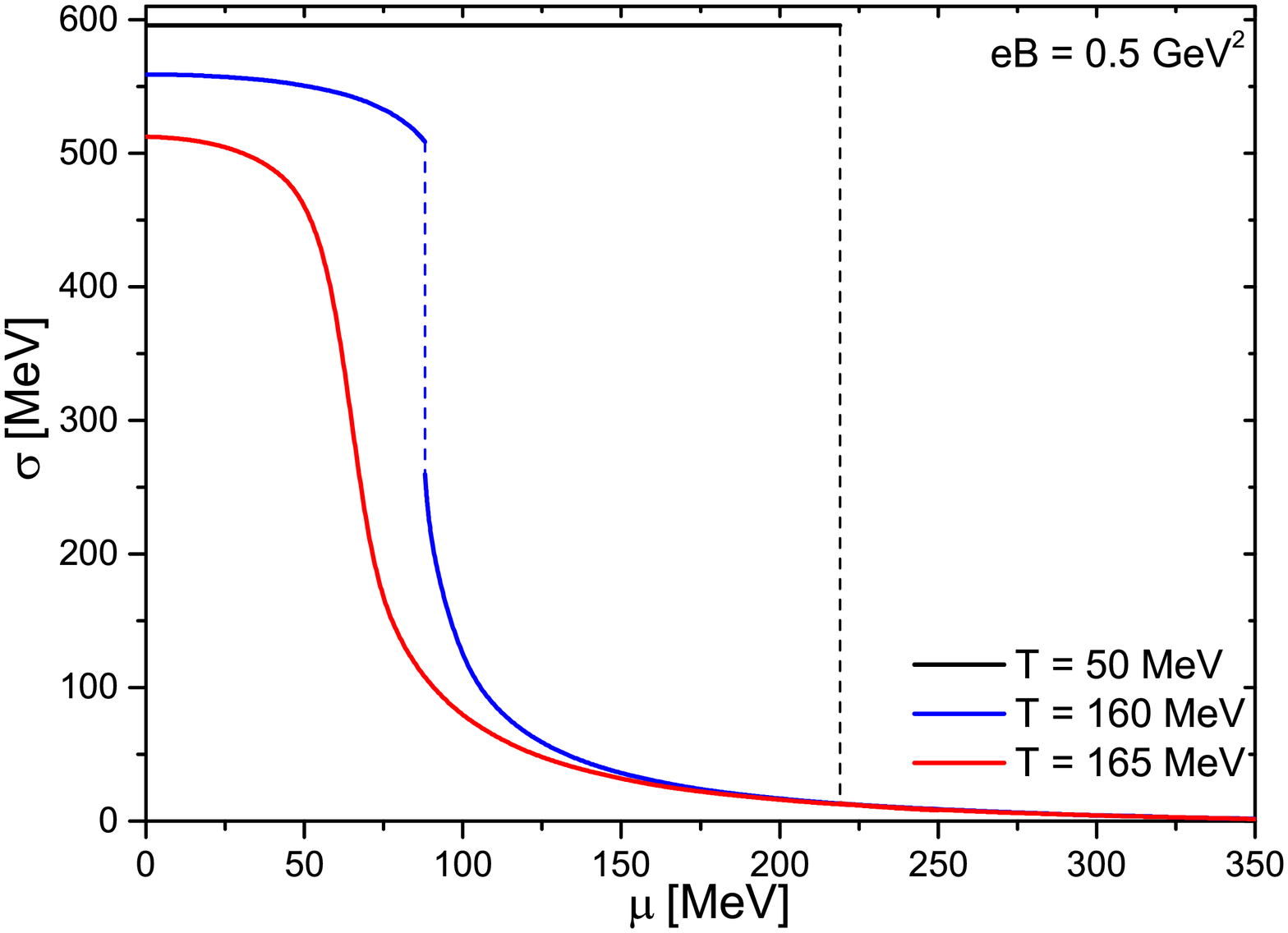}
\end{tabular}
\caption{ Behavior of $\bar{\sigma}$ as a function of $\mu$ for different values of the temperature and magnetic field.} 
\label{fig1}
\end{figure}

We numerically calculate the values of the scalar fields $\bar{\sigma}$ and $\Phi$ that satisfy the coupled gap-equations Eq.~(\ref{gap}). In Fig.~\ref{fig1}, we show the behavior of $\bar{\sigma}$ as a function of chemical potential $\mu$ for two representative values of magnetic field as $eB=0$ (upper panel) and $eB=0.5$~GeV$^{2}$ (lower panel), and different values of temperature, $T=50,160,165$ MeV. For both values of magnetic field, in the temperature range from $\sim$ 50 to 160~MeV we can observe that $\bar{\sigma}$ has a discontinuity at $\mu=\mu_{c}$ that represents a first order transition, in which at low values of chemical potential, $\mu<\mu_{c}$, the system is in a chiral symmetry broken phase where quarks acquire dynamical mass, and for values $\mu>\mu_{c}$ the system partially recovers chiral symmetry. By increasing the temperature, the discontinuity of the mean field $\bar{\sigma}$ related to the first order transition transforms into a crossover-like transition. This can be observed in the lower panel of Fig.~\ref{fig1}, by looking at the curve of $\bar{\sigma}$ vs. $\mu$ (red line) for $eB=0.5$~GeV$^{2}$ and $T=165$ MeV. For that temperature, for $eB=0$ (upper panel) the smooth transition is still not present.                    

\begin{figure}[htb]
\includegraphics[scale=0.31]{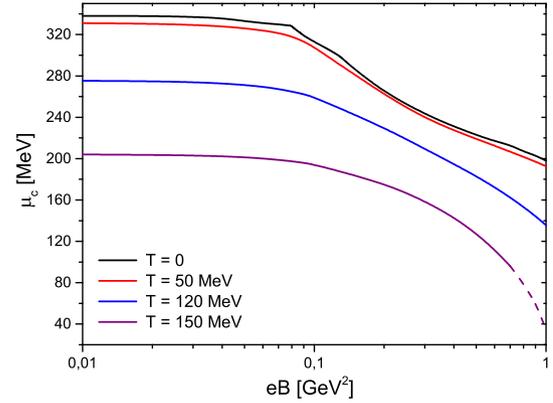} 
\caption{Phase diagram in $\mu_{c}-eB$ plane for different values of temperature. Solid and dashed lines represent first order and crossover phase transitions, respectively.} 
\label{fig2}
\end{figure}

In Fig.~\ref{fig2} we plot the phase diagrams in $\mu_{c}-eB$ plane for different values of temperature, namely $T=0,50,120,150$~MeV. The phase diagram corresponding to $T=0$ was already obtained in a previous work \citep{Ferraris_2021}. From the phase diagrams corresponding to $T=50$~MeV, $T=120$~MeV and $T=150$~MeV, we can observe a similar qualitative behavior to $T=0$. The critical chemical potential $\mu_{c}$ shows a weak dependence on the magnetic field for values lower than $eB\sim 0.06$~GeV$^{2}$. On the other hand, for larger values, the critical chemical potential $\mu_{c}$ becomes a decreasing function of the magnetic field. This decreasing behavior of $\mu_{c}$ as a function of $eB$ is known as \textit{inverse magnetic catalysis} (IMC) \citep{allen2013quark}. It is important to notice that for all phase diagrams, even for the one calculated at $T=0$, the solid lines represent a first order transition. The lines divide the diagrams into two regions, one corresponding to a chiral symmetry broken phase (lower values of $\mu_{c}$, massive phase) and another one in which chiral symmetry is partially restored (higher values of $\mu_{c}$). In particular, for $T=150$~MeV one finds a crossover transition in the region of strong magnetic fields, $eB\sim0.8$~GeV$^{2}$.

\begin{figure}[htb]
\includegraphics[scale=0.31]{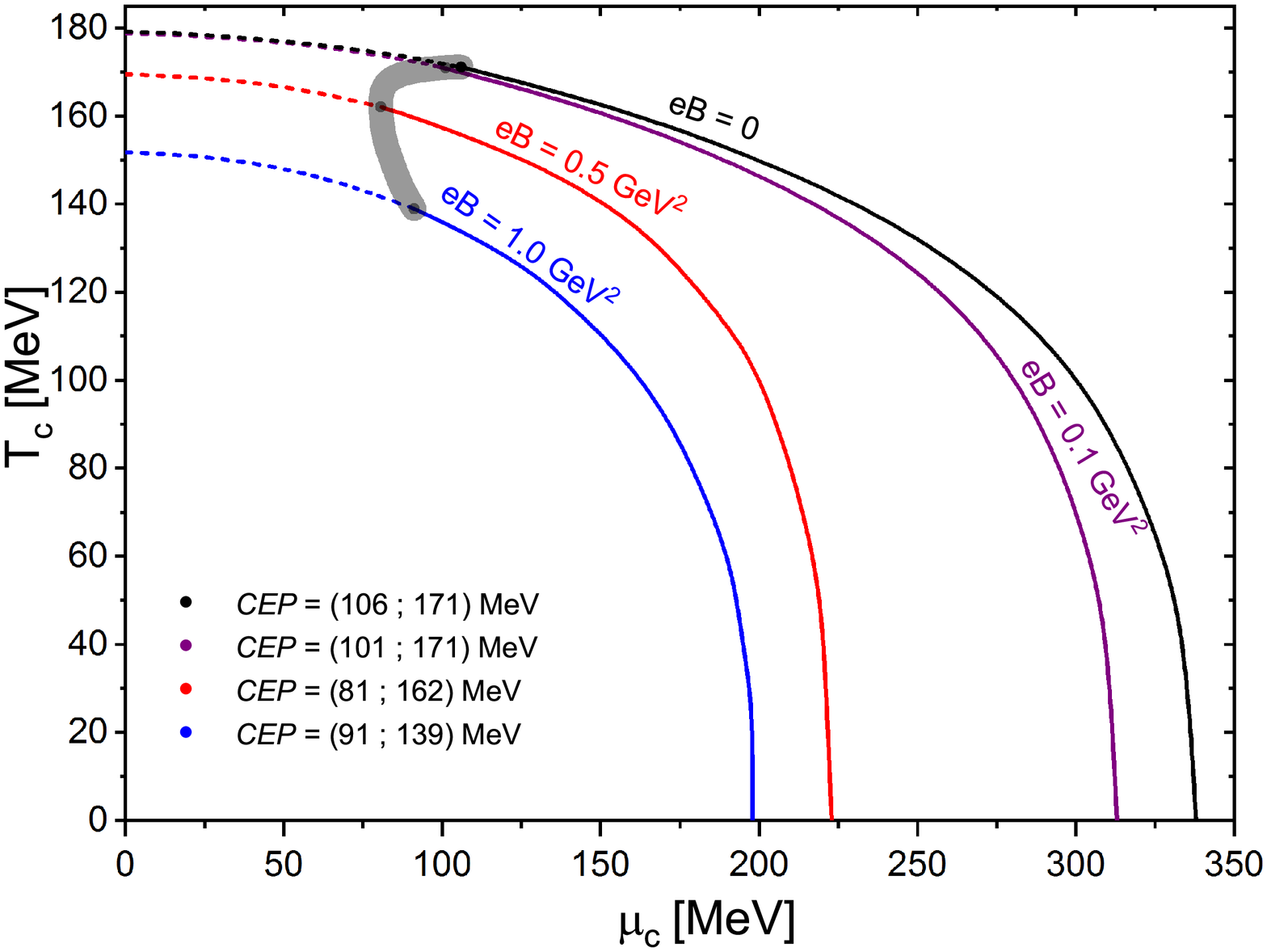} 
\caption{Phase diagram in $T_{c}-\mu_{c}$ plane for different values of magnetic fields. Solid and dashed lines represent first order and crossover phase transitions, respectively.} 
\label{fig3}
\end{figure}

 In Fig.~\ref{fig3} we show different phase diagrams in $T_{c}-\mu_{c}$ plane calculated at the representative magnetic field values $eB=0,~0.1,~0.5$~GeV$^{2}$ and $eB=1.0$~GeV$^{2}$. As a general feature, for all phase diagrams, we observe a crossover transition in the region of high critical temperature $T_{c}$ and low critical chemical potential $\mu_{c}$ (dashed lines). The crossover transitions were calculated considering the maxima of the \textit{chiral susceptibility}, defined as the derivative $\chi_{ch}=-\partial\left[ \left(\braket{\bar{u}u}_{B,T,\mu}^{\rm reg} +  \braket{\bar{d}d}_{B,T,\mu}^{\rm reg} \right)/2 \right]/\partial m_c$. By increasing the chemical potential, the crossover transition becomes a first order transition (solid lines) at the \textit{critical end-point} CEP. The wide gray line shown in Fig.~\ref{fig3} indicates the location of critical end-points for the phase diagrams corresponding to the range $eB = 0$ to $1\ {\rm GeV}^2$. 
 Note that the crossover transition lines (dashed lines) are found to be approximately overlapped for low magnetic field values $eB<0.1$~GeV$^{2}$; the locations of the CEP values in this magnetic field range are also coincident. 

 As shown in Fig.~\ref{fig2}, for low values of the chemical potential we observe that the critical temperature decreases as the magnetic field increases (a manifestation of inverse magnetic catalysis). Finally, for low fixed temperatures, the increasing magnetic fields anticipate the chiral symmetry restoration via a first order phase transition.

 \section{Conclusions}

 In the present work we have studied the phase diagram of magnetized quark matter, at both finite temperature and chemical potential, for different values of the magnetic field. We have considered a two-flavor non-local version of the NJL model, with a Gaussian form factor. We have considered as well the interaction between the quarks and the Polyakov loop and introduced a polynomial PL potential.

We have worked in the mean-field approximation, obtaining numerically the effective masses and the Polyakov loop by solving self-consistently the gap equations for given values of temperature, chemical potential, and magnetic field.
The model parameters have been taken from Refs.~\citep{dumm2017strong} and \citep{Ferraris_2021}, where temperature and chemical potential effects were considered separately. For this parameterization, our results show that that the $T_{c} -\mu_{c}$ phase diagrams for different values of the magnetic field are qualitatively similar to each other. 
For all magnetic field strengths considered in the present work it is seen that there is always a critical end-point (CEP) that connects a first-order transition with a smooth crossover.
As in the case of non-magnetized quark matter, in the region of high temperatures and low chemical potentials the system undergoes a smooth crossover transition to the deconfined phase where the chiral symmetry is (partially) restored. By increasing the chemical potential, one reaches a critical end point beyond which the transition becomes of first order. For the considered range of values of the magnetic field, it is found that the CEP lies in a region given by $80$~MeV~$< \mu_c < 106$~MeV and $139$~MeV~$< T < 172$~MeV.


\begin{thebibliography}{}

\bibitem [\protect \citeauthoryear {%
Allen%
\ \BBA {} Scoccola%
}{%
Allen%
\ \BBA {} Scoccola%
}{%
{\protect \APACyear {2013}}%
}]{%
allen2013quark}
\APACinsertmetastar {%
allen2013quark}%
\begin{APACrefauthors}%
Allen, P\BPBI G.%
\BCBT {}\ \BBA {} Scoccola, N\BPBI N.%
\end{APACrefauthors}%
\unskip\
\newblock
\APACrefYearMonthDay{2013}{}{},
\newblock
\unskip
\newblock
\APACjournalVolNumPages{Physical Review D}{88}{9}{094005}.
\PrintBackRefs{\CurrentBib}

\bibitem [\protect \citeauthoryear {%
Bloch%
}{%
Bloch%
}{%
{\protect \APACyear {1952}}%
}]{%
bloch1952field}
\APACinsertmetastar {%
bloch1952field}%
\begin{APACrefauthors}%
Bloch, C.%
\end{APACrefauthors}%
\unskip\
\newblock
\APACrefYearMonthDay{1952}{}{},
\newblock
\unskip
\newblock
\APACjournalVolNumPages{Kgl. Danske Videnskab. Selskab, Mat. fys.
  Medd.}{27}{}{}.
\PrintBackRefs{\CurrentBib}

\bibitem [\protect \citeauthoryear {%
Deng%
\ \BBA {} Huang%
}{%
Deng%
\ \BBA {} Huang%
}{%
{\protect \APACyear {2012}}%
}]{%
Deng:2012pc}
\APACinsertmetastar {%
Deng:2012pc}%
\begin{APACrefauthors}%
Deng, W\BHBI T.%
\BCBT {}\ \BBA {} Huang, X\BHBI G.%
\end{APACrefauthors}%
\unskip\
\newblock
\APACrefYearMonthDay{2012}{}{},
\newblock
\unskip
\newblock
\APACjournalVolNumPages{Phys. Rev. C}{85}{}{044907}.
\PrintBackRefs{\CurrentBib}

\bibitem [\protect \citeauthoryear {%
Dumitru%
, Pisarski%
\BCBL {}\ \BBA {} Zschiesche%
}{%
Dumitru%
\ \protect \BOthers {.}}{%
{\protect \APACyear {2005}}%
}]{%
dumitru2005dense}
\APACinsertmetastar {%
dumitru2005dense}%
\begin{APACrefauthors}%
Dumitru, A.%
, Pisarski, R\BPBI D.%
\BCBL {}\ \BBA {} Zschiesche, D.%
\end{APACrefauthors}%
\unskip\
\newblock
\APACrefYearMonthDay{2005}{}{},
\newblock
\unskip
\newblock
\APACjournalVolNumPages{Physical Review D}{72}{6}{065008}.
\PrintBackRefs{\CurrentBib}

\bibitem [\protect \citeauthoryear {%
Dumm%
, Grunfeld%
\BCBL {}\ \BBA {} Scoccola%
}{%
Dumm%
\ \protect \BOthers {.}}{%
{\protect \APACyear {2006}}%
}]{%
dumm2006covariant}
\APACinsertmetastar {%
dumm2006covariant}%
\begin{APACrefauthors}%
Dumm, D\BPBI G.%
, Grunfeld, A.%
\BCBL {}\ \BBA {} Scoccola, N.%
\end{APACrefauthors}%
\unskip\
\newblock
\APACrefYearMonthDay{2006}{}{},
\newblock
\unskip
\newblock
\APACjournalVolNumPages{Physical Review D}{74}{5}{054026}.
\PrintBackRefs{\CurrentBib}

\bibitem [\protect \citeauthoryear {%
Dumm%
\ \BBA {} Scoccola%
}{%
Dumm%
\ \BBA {} Scoccola%
}{%
{\protect \APACyear {2005}}%
}]{%
dumm2005characteristics}
\APACinsertmetastar {%
dumm2005characteristics}%
\begin{APACrefauthors}%
Dumm, D\BPBI G.%
\BCBT {}\ \BBA {} Scoccola, N\BPBI N.%
\end{APACrefauthors}%
\unskip\
\newblock
\APACrefYearMonthDay{2005}{}{},
\newblock
\unskip
\newblock
\APACjournalVolNumPages{Physical Review C}{72}{1}{014909}.
\PrintBackRefs{\CurrentBib}

\bibitem [\protect \citeauthoryear {%
Dumm%
, Villafa{\~n}e%
, Noguera%
, Pagura%
\BCBL {}\ \BBA {} Scoccola%
}{%
Dumm%
\ \protect \BOthers {.}}{%
{\protect \APACyear {2017}}%
}]{%
dumm2017strong}
\APACinsertmetastar {%
dumm2017strong}%
\begin{APACrefauthors}%
Dumm, D\BPBI G.%
, Villafa{\~n}e, M\BPBI I.%
, Noguera, S.%
, Pagura, V\BPBI P.%
\BCBL {}\ \BBA {} Scoccola, N\BPBI N.%
\end{APACrefauthors}%
\unskip\
\newblock
\APACrefYearMonthDay{2017}{}{},
\newblock
\unskip
\newblock
\APACjournalVolNumPages{Physical Review D}{96}{11}{114012}.
\PrintBackRefs{\CurrentBib}

\bibitem [\protect \citeauthoryear {%
{Duncan}%
\ \BBA {} {Thompson}%
}{%
{Duncan}%
\ \BBA {} {Thompson}%
}{%
{\protect \APACyear {1992}}%
}]{%
1992ApJ...392L...9D}
\APACinsertmetastar {%
1992ApJ...392L...9D}%
\begin{APACrefauthors}%
{Duncan}, R\BPBI C.%
\BCBT {}\ \BBA {} {Thompson}, C.%
\end{APACrefauthors}%
\unskip\
\newblock
\APACrefYearMonthDay{1992}{}{},
\newblock
\unskip
\newblock
\APACjournalVolNumPages{\apjl}{392}{}{L9}.
\PrintBackRefs{\CurrentBib}

\bibitem [\protect \citeauthoryear {%
Ferraris%
, Dumm%
, Grunfeld%
\BCBL {}\ \BBA {} Scoccola%
}{%
Ferraris%
\ \protect \BOthers {.}}{%
{\protect \APACyear {2021}}%
}]{%
Ferraris_2021}
\APACinsertmetastar {%
Ferraris_2021}%
\begin{APACrefauthors}%
Ferraris, S\BPBI A.%
, Dumm, D\BPBI G.%
, Grunfeld, A\BPBI G.%
\BCBL {}\ \BBA {} Scoccola, N\BPBI N.%
\end{APACrefauthors}%
\unskip\
\newblock
\APACrefYearMonthDay{2021}{}{},
\newblock
\unskip
\newblock
\APACjournalVolNumPages{The European Physical Journal A}{57}{4}{}.
\PrintBackRefs{\CurrentBib}

\bibitem [\protect \citeauthoryear {%
G{\'o}mez~Dumm%
, Noguera%
\BCBL {}\ \BBA {} Scoccola%
}{%
G{\'o}mez~Dumm%
\ \protect \BOthers {.}}{%
{\protect \APACyear {2011}}%
}]{%
gomez2011pion}
\APACinsertmetastar {%
gomez2011pion}%
\begin{APACrefauthors}%
G{\'o}mez~Dumm, D\BPBI A.%
, Noguera, S.%
\BCBL {}\ \BBA {} Scoccola, N\BPBI N.%
\end{APACrefauthors}%
\unskip\
\newblock
\APACrefYearMonthDay{2011}{}{},
\newblock
\unskip
\newblock
\APACjournalVolNumPages{Physics Letters B}{698}{}{}.
\PrintBackRefs{\CurrentBib}

\bibitem [\protect \citeauthoryear {%
Noguera%
\ \BBA {} Scoccola%
}{%
Noguera%
\ \BBA {} Scoccola%
}{%
{\protect \APACyear {2008}}%
}]{%
noguera2008nonlocal}
\APACinsertmetastar {%
noguera2008nonlocal}%
\begin{APACrefauthors}%
Noguera, S.%
\BCBT {}\ \BBA {} Scoccola, N.%
\end{APACrefauthors}%
\unskip\
\newblock
\APACrefYearMonthDay{2008}{}{},
\newblock
\unskip
\newblock
\APACjournalVolNumPages{Physical Review D}{78}{11}{114002}.
\PrintBackRefs{\CurrentBib}

\bibitem [\protect \citeauthoryear {%
Ratti%
, Thaler%
\BCBL {}\ \BBA {} Weise%
}{%
Ratti%
\ \protect \BOthers {.}}{%
{\protect \APACyear {2006}}%
}]{%
ratti2006phases}
\APACinsertmetastar {%
ratti2006phases}%
\begin{APACrefauthors}%
Ratti, C.%
, Thaler, M\BPBI A.%
\BCBL {}\ \BBA {} Weise, W.%
\end{APACrefauthors}%
\unskip\
\newblock
\APACrefYearMonthDay{2006}{}{},
\newblock
\unskip
\newblock
\APACjournalVolNumPages{Physical Review D}{73}{1}{014019}.
\PrintBackRefs{\CurrentBib}

\bibitem [\protect \citeauthoryear {%
Ritus%
}{%
Ritus%
}{%
{\protect \APACyear {1978}}%
}]{%
ritus1978method}
\APACinsertmetastar {%
ritus1978method}%
\begin{APACrefauthors}%
Ritus, V.%
\end{APACrefauthors}%
\unskip\
\newblock
\APACrefYearMonthDay{1978}{}{},
\newblock
\APACrefbtitle {Method of eigenfunctions and mass operator in quantum
  electrodynamics of a constant field} {Method of eigenfunctions and mass
  operator in quantum electrodynamics of a constant field}\
  \APACbVolEdTR{}{\BTR{}}.
\newblock
\APACaddressInstitution{}{CM-P00067532}.
\PrintBackRefs{\CurrentBib}

\bibitem [\protect \citeauthoryear {%
Scavenius%
, Dumitru%
\BCBL {}\ \BBA {} Lenaghan%
}{%
Scavenius%
\ \protect \BOthers {.}}{%
{\protect \APACyear {2002}}%
}]{%
scavenius2002k}
\APACinsertmetastar {%
scavenius2002k}%
\begin{APACrefauthors}%
Scavenius, O.%
, Dumitru, A.%
\BCBL {}\ \BBA {} Lenaghan, J.%
\end{APACrefauthors}%
\unskip\
\newblock
\APACrefYearMonthDay{2002}{}{},
\newblock
\unskip
\newblock
\APACjournalVolNumPages{Physical Review C}{66}{3}{034903}.
\PrintBackRefs{\CurrentBib}

\bibitem [\protect \citeauthoryear {%
Vachaspati%
}{%
Vachaspati%
}{%
{\protect \APACyear {1991}}%
}]{%
Vachaspati:1991nm}
\APACinsertmetastar {%
Vachaspati:1991nm}%
\begin{APACrefauthors}%
Vachaspati, T.%
\end{APACrefauthors}%
\unskip\
\newblock
\APACrefYearMonthDay{1991}{}{},
\newblock
\unskip
\newblock
\APACjournalVolNumPages{Phys. Lett. B}{265}{}{258--261}.
\PrintBackRefs{\CurrentBib}

\end{thebibliography}

\end{document}